# Using a Conditional Generative Adversarial Network to Control the Statistical Characteristics of Generated Images for IACT Data Analysis

**Julia Dubenskaya**[a,*], **Alexander Kryukov**[a], **Andrey Demichev**[a], **Stanislav Polyakov**[a], **Elizaveta Gres**[b], **Anna Vlaskina**[c]

[a] *Skobeltsyn Institute of Nuclear Physics, Lomonosov Moscow State University,*
  *1(2), Leninskie gory, GSP-1, Moscow 119991, Russia*

[b] *Applied Physics Institute of Irkutsk State University,*
  *20, Gagarin boulevard, Irkutsk, 664003, Russia*

[c] *Lomonosov Moscow State University,*
  *1(2), Leninskie gory, GSP-1, Moscow 119991, Russia*

  *E-mail:* jdubenskaya@gmail.com

Generative adversarial networks are a promising tool for image generation in the astronomy domain. Of particular interest are conditional generative adversarial networks (cGANs), which allow you to divide images into several classes according to the value of some property of the image, and then specify the required class when generating new images. In the case of images from Imaging Atmospheric Cherenkov Telescopes (IACTs), an important property is the total brightness of all image pixels (image size), which is in direct correlation with the energy of primary particles. We used a cGAN technique to generate images similar to whose obtained in the TAIGA-IACT experiment. As a training set, we used a set of two-dimensional images generated using the TAIGA Monte Carlo simulation software. We artificiallly divided the training set into 10 classes, sorting images by size and defining the boundaries of the classes so that the same number of images fall into each class. These classes were used while training our network. The paper shows that for each class, the size distribution of the generated images is close to normal with the mean value located approximately in the middle of the corresponding class. We also show that for the generated images, the total image size distribution obtained by summing the distributions over all classes is close to the original distribution of the training set. The results obtained will be useful for more accurate generation of realistic synthetic images similar to the ones taken by IACTs.

[*]Speaker

# 1. Introduction

In recent years, due to the increase in the power of computer resources, generative adversarial networks or GANs [1] achieved amazing results in many tasks of image processing and generation. In particular, GANs are increasingly being used in various fields of astrophysics, both for generating synthetic images that are difficult to distinguish from real ones, and for processing real images such as ones taken by ground and space telescopes.

GANs were applied to generate astronomical images of galaxies [2] as well as solar images [3], to improve noisy astrophysical images [4], to perform simulations of cosmic web [5], to produce realistic synthetic images from the Hubble Space Telescope [6], and to retrieve exoplanetary atmospheres [7]. In the task of generating images, in most cases GANs are used to bypass detailed direct simulations of the underlying physical processes that are usually computationally intensive, time consuming, and therefore expensive.

In our previous work [8, 9, 10], we used GANs to generate images similar to those taken by Imagine Atmospheric Cherenkov telescopes (IACTs) in the TAIGA-IACT experiment [11]. IACTs capture the Cherenkov light emitted when an extensive air shower (EAS) appears after a high-energy particle hits the atmosphere. During operation, each IACT records a set of images of the air shower against the background light of the night sky. Each image is a set of pixels, and the pixel values are given in photoelectrons.

In the TAIGA-IACT experiment, simulation images are used for comparisons to real data and to estimate the performance of the detector setup [12]. Traditionally, event images for the TAIGA-IACT project are modeled using special software for realistic Monte-Carlo simulations. First, the shower itself is simulated using the CORSIKA toolkit [13], that performs detailed direct simulation of EAS evolution. The response of the IACT system is simulated using the special software [14] that performs a full ray tracing of the Cherenkov photons through the telescopes' optics. These programs, when used together, give very accurate results, but they work quite slowly, specifically, they generate an average of 1000 images per hour. For some analysis purposes such as synthetic minority oversampling [15], such simulation accuracy is redundant, so that less complex and more efficient generation methods can be used. Previously, we have shown that the adoption of the GAN technique accelerated the generation of images by 1500 times compared to the direct simulation method [8], while the quality of the generated images remained very high, and most of the generated images were statistically indistinguishable from the images of the training set.

However, it should be noted that often for scientific applications, in addition to the quality of each individual image, it is also very important to reproduce the statistical characteristics of the reference data (real or model) in the sample of the generated images. When generating IACT images, it is important to reproduce the distribution of images over the energy of the primary particle. Simulated data that meet this requirement are not statistically different from actually observed data. The use of such simulated data leads to a more accurate detector tuning and, accordingly, to a more accurate identification of the observed events.

The images generated using GAN do not contain any information concerning the primary particle, and restoring the energy of the original particle from the generated images is a separate issue. That's why as a first approximation, instead of energy, it is convenient to use the total brightness of all image pixels called image size. The image size is the total sum of all image pixel values in photoelectrons, and it's great advantage is that it can be easily calculated for any arbitrary image. At the same time, this value is in direct correlation with the energy of a primary particle [16].

In our previous work we have shown that the GAN was unable to fully reproduce the statistical parameters of the training set as a whole [9], and when comparing the image size distribution of the training set with one of the GAN output, we found a significant discrepancy. While the shapes of the distributions were generally similar, the frequency of occurrence of rare events (ones with very low or very high image sizes) was much less for the GAN output. Solving the problem using classical GAN was difficult because this type of network does not provide means for managing the properties of the generated images. The situation is different for conditional generative adversarial networks or cGANs [17] that are an extension of GANs for conditional sample generation. cGAN gives you control over the parameters of the generated data. Namely, cGAN allows you to divide images into multiple classes according to the value of some property of the image, and then specify the required class when generating each new image. Implementing cGAN helped us to significantly improve the image size distribution of the generated sample and bring it closer to the distribution of the training set.

## 2. Image size distribution of the input sample

As a training set, we used a sample of two-dimensional images obtained using TAIGA Monte Carlo simulation software. This software performs direct simulations of EAS evolution [13] and IACT system response [14] and generates a set of images similar to those actually detected. The training set contains 35000 gamma event images. By a gamma event we mean a detected Cherenkov light that occurs in an EAS induced by a high-energy gamma ray. During the training process, the training set was fed to the input of our neural networks.

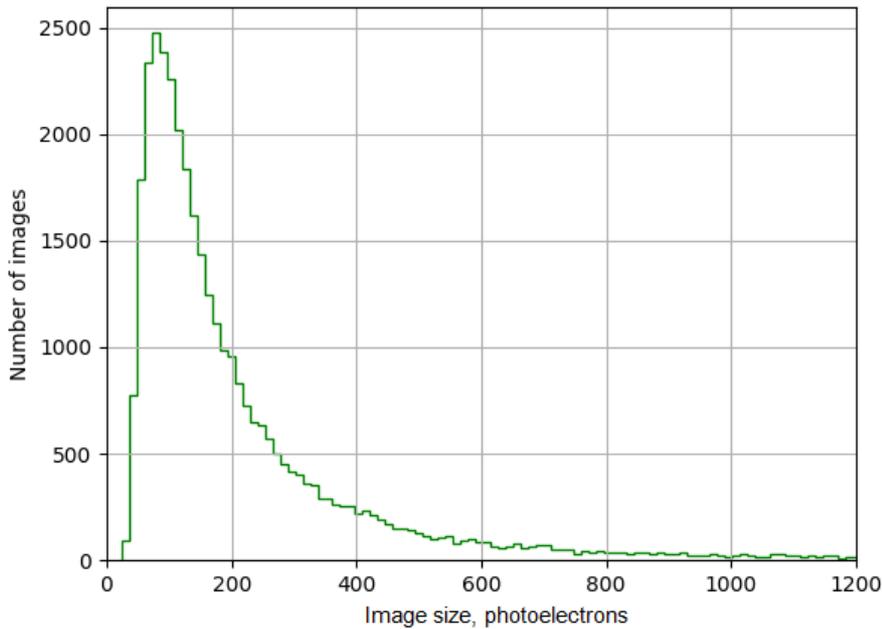

**Figure 1.** Image size distribution for the training set

If we look at the image size distribution (see Figure 1) for the training set, we can see that this distribution is very uneven and asymmetric. Namely, the distribution has one maximum for a relatively small image size value, while the number of images with very small and very large image sizes is not too big and decreases with distance from the maximum.

First, we generated images using classical GAN, and compared the image size distributions for the training set and for the GAN-generated sample. We found that these two distributions differ significantly, with the chi-square test statistic being about 4900. The corresponding graphs are shown in Figure 2.

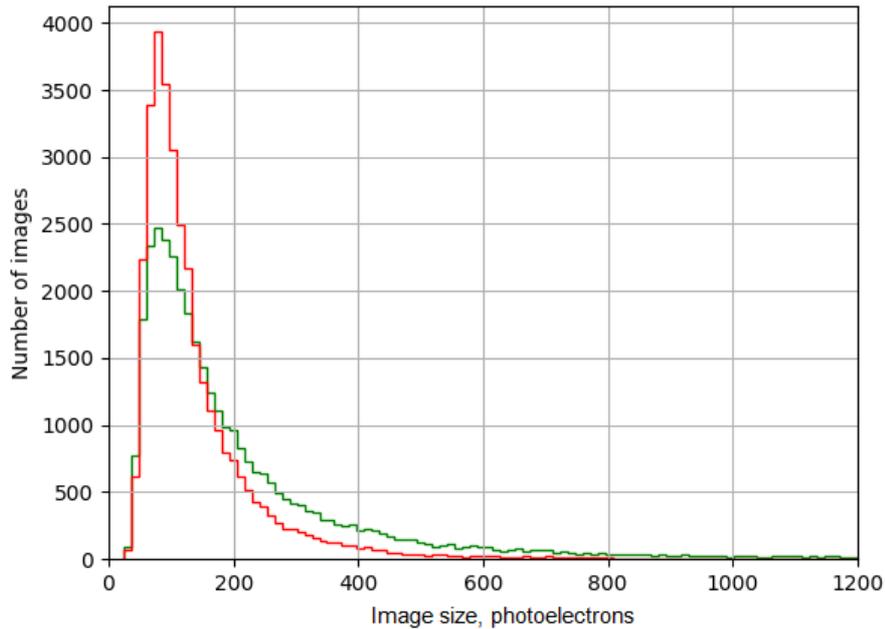

**Figure 2.** Image size distribution for the training set (green) and for the GAN-generated sample (red)

As you can see from the graph, GAN mainly generates images from the range that most input images fall within. This behavior can be explained as follows. The neural network is trained by examples. There is an uneven number of examples with different image sizes in the training set. This leads to the fact that the neural network is better trained to generate images similar to those that appear more often in the training set. Therefore, the image size distributions for the training set and for the GAN-generated sample are very different. To address this issue, we propose to use a cGAN instead of a classical GAN to control the image size values of the generated images.

## 3. Conditional Generative Adversarial Network for TAIGA-IACT images

### 3.1 cGAN and artificial classes for IACT images

cGAN is a modification of a classical GAN that involves the conditional generation of images by a generator model. Like any GAN, a cGAN consists of two neural networks: a generator and a discriminator that are trained together on real images in a zero-sum game.

To train a cGAN, you first need to manually prepare a set of training images labeled according to the following rules:
- Images are separated into several mutually exclusive classes based on the value of some image property.
- Each image is assigned a label corresponding to its class number.
- The number of classes (N) is a positive integer.
- There are no images without a class label.
- The training set contains images from each class.

Then a cGAN is trained on the set prepared in this way. A set of labeled images is passed to the discriminator. The discriminator is trained to distinguish whether an image is real or fake given a class label. The generator attempts to create fake images of the requested class that the discriminator would consider real.

A cGAN do very well when images from different classes differ significantly. As a striking example, consider images of handwritten digits. Each digit can always be attributed to a particular class. Totally there are 10 classes, and there is no transition digits between classes, for example between 7 and 8. In this case, a cGAN easily learns to generate images of the specified class. For images from the TAIGA Cherenkov telescopes the situation is quite different. Given that we want to control the image size, we need to separate images into classes based on the image size values, which can be any real number. Furthermore, the image size distribution is uneven and asymmetric with a single maximum, so we can only separate classes artificially. Keeping in mind that the uneven number of examples in the training set leads to a distortion of the output distribution, we applied an artificial separation of the images of the training set into 10 classes, sorting the images by size and defining the boundaries of the classes so that the same number of images fall into each class. Thus, for our training set, we got 3500 images in each class. We hoped that with this approach, we would not lose rare events, and the training would become more balanced and stable.

The input image size distribution and class boundaries are shown in Figure 3.

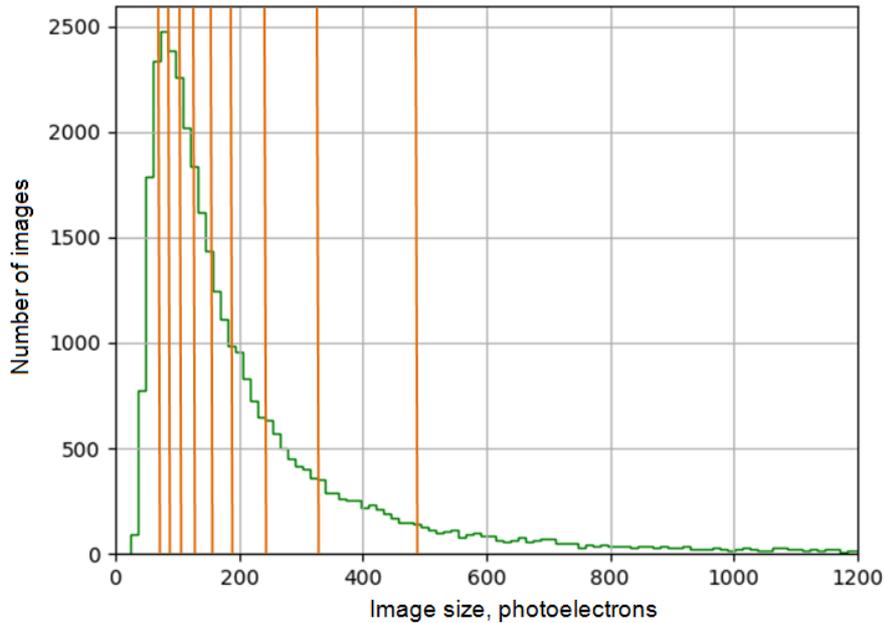

**Figure 3.** Image size distribution for the training set (green) and class boundaries (orange)

Thus, each image of the training set was labeled with a class number according to its size. Additionally, as with a classical GAN, for a cGAN, before training each image must be transformed as follows. The TAIGA-IACT telescope cameras consist of arrays of photomultipliers arranged in a hexagonal grid, and each photomultiplier produces one image pixel. Accordingly, the IACT images are also hexagonal. However, current high performance cGAN implementations are designed for square grids. Therefore we had to convert the hexagonal images to square ones by transition to an oblique coordinate system. As a result, we got images of 32 by 32 pixels. Then, the pixel values were scaled to the range [0, 1] to match the output of the generator model. Unlike many other image types, the pixel value of which lies in a fixed range (for example, from 0 to 256), the pixel value in photoelectrons is theoretically unlimited from above. Therefore, when moving to the range [0, 1], we calculate the maximum pixel value for our training set, which we keep in mind and use later to reverse convert the generated images. This is how we get a square grayscale image which we feed to the discriminator input.

An example of the original image and the image after preprocessing is shown in Figure 4.

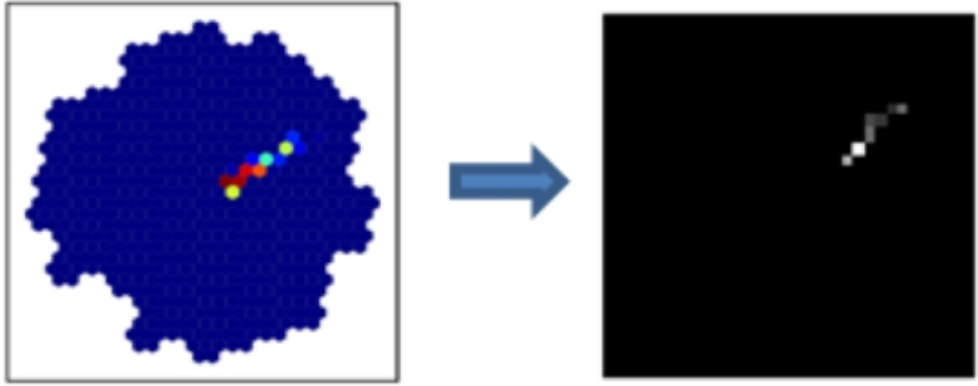

**Figure 4.** An example of an input image from the training set (the original image and the same image after preprocessing)

### 3.2 Proposed cGAN architecture

Since a cGAN consists of a discriminator and a generator, we next describe the architecture of each of these networks in turn.

The discriminator architecture is shown in Figure 5. The discriminator is a small network consisting of a convolutional layer with 3x3 filters followed by a dense (fully connected) layer with 64 neurons in it. The convolutional layers use a leaky ReLU function with alpha=0.2 as the activation function. The output layer uses a sigmoid as the activation function.

The architecture of the generator is shown in Figure 6. The generator takes a random vector and a class label as input, and then uses transpose convolution to upsample until it gets an image with the desired number of pixels. All layers except the output layer use 3x3 filters and a leaky ReLU function with alpha=0.2 as the activation function. The output layer has one 5x5 filter and uses a hyperbolic tangent as the activation function.

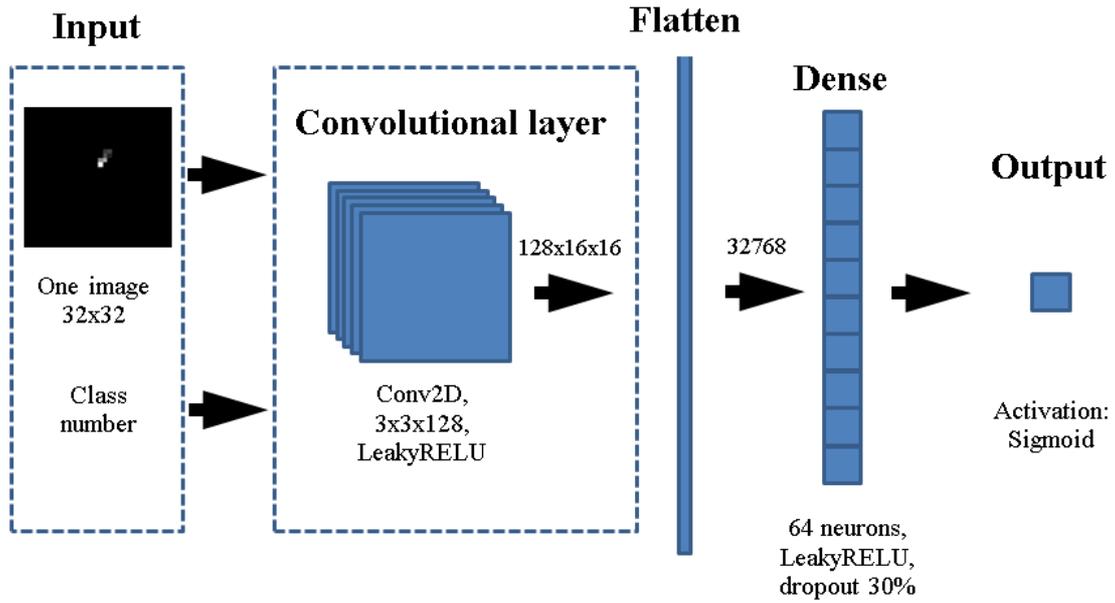

**Figure 5.** Architecture of the discriminator for TAIGA-IACT images

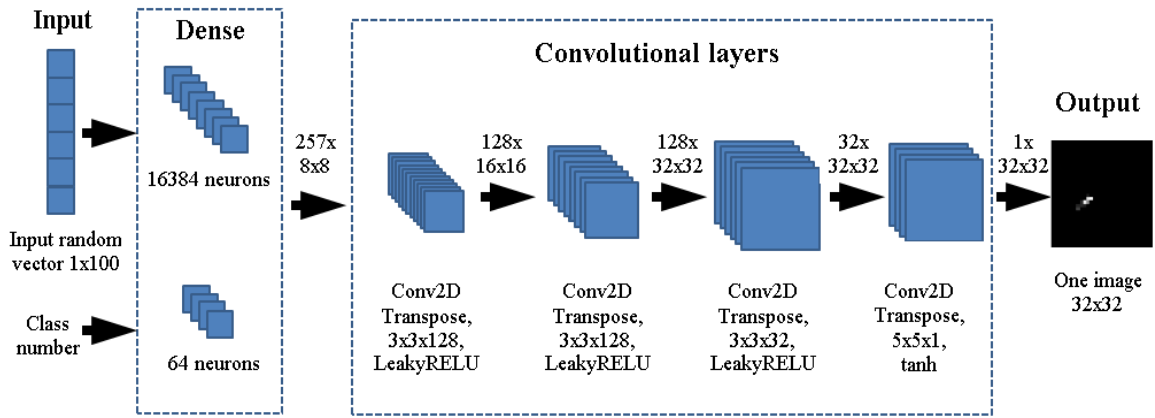

**Figure 6.** Architecture of the generator for TAIGA-IACT images

### 3.3 Training results

We used the above mentioned ten artificial classes while training our cGAN. We have implemented the network with the proposed architecture using the TensorFlow [18] software package. Network training at the GPU Tesla P100 with a batch size of 128 images and 500 epochs took about 8 hours. After training, generation of 10000 events (of any class) takes about 12 seconds.

After training, we used the generator to create images of each class. We generated 3500 images per each class, and this number was exactly equal to the number of images in each of the training set classes. The mean image size and a standard deviation for each class according to the class boundaries are shown in Figure 7.

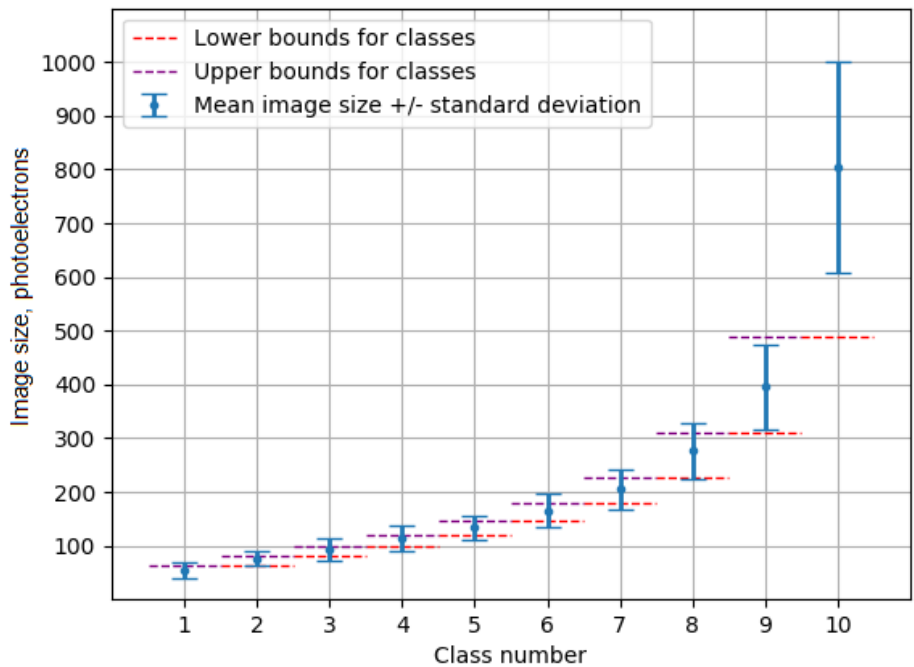

**Figure 7.** A mean image size (blue dots) and a standard deviation for each class

We have got that for each class, the image size distribution is close to normal with an average value located approximately in the middle of the corresponding class. This can be seen in more detail on the example of one of the classes. Here is the distribution for the class number two (see Figure 8).

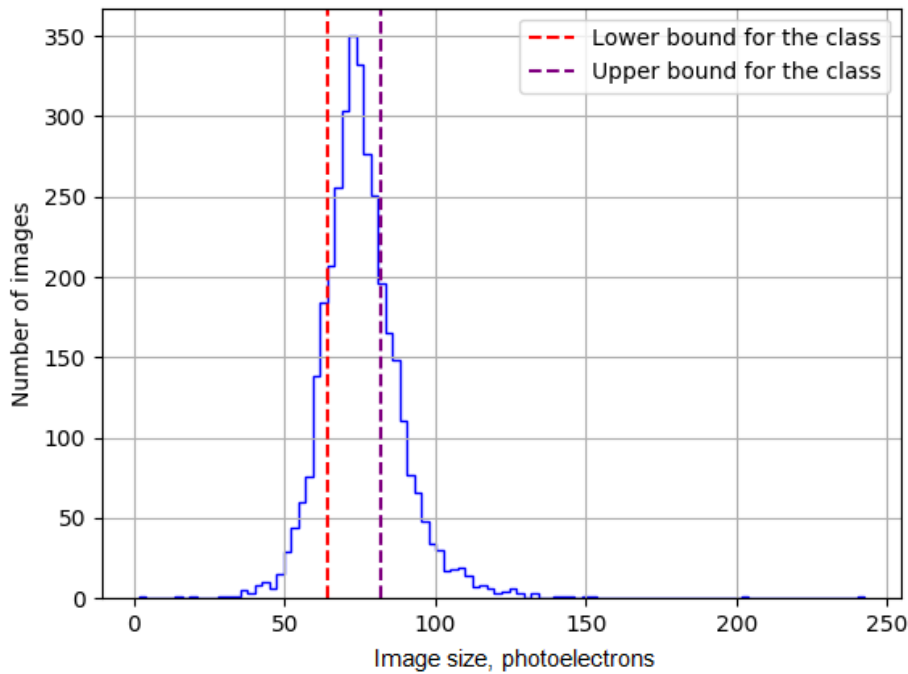

**Figure 8.** Image size distribution for the cGAN output for class #2 (blue), and lower (red) and upper (purple) class boundaries

As you can see on the graph, the image size distribution for the generated images is close to normal with the mean located approximately halfway between the class boundaries. Also you can see that the distribution goes beyond the class boundaries. Going beyond the boundaries is not very good when we want to generate an image of a particular class, but this feature is very useful for the total distribution. The total image size distribution obtained by summing the distributions over all classes is shown in Figure 9.

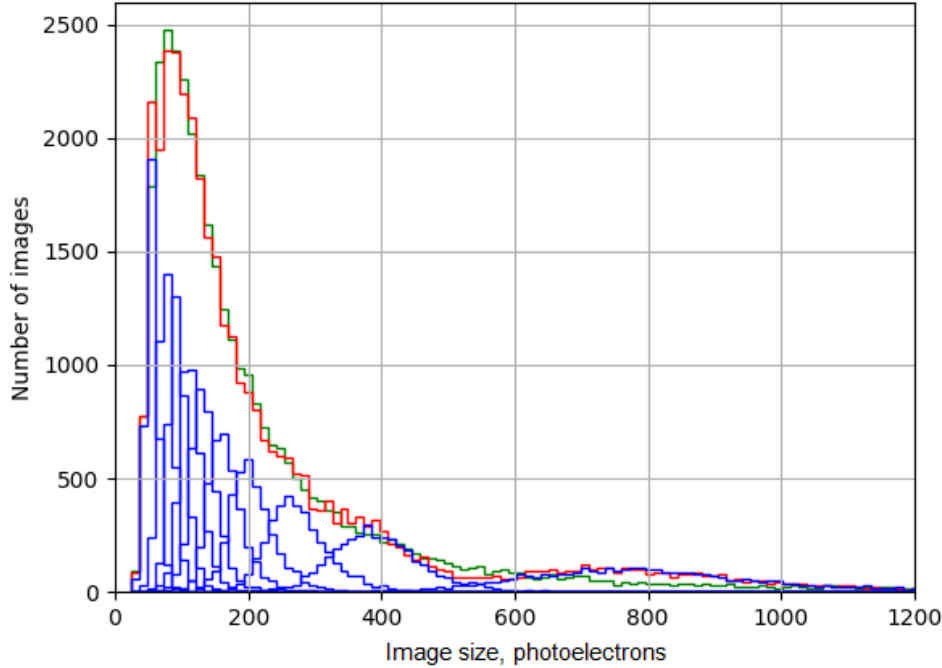

**Figure 9.** Image size distribution of the training set (green), image size distribution for the ten cGAN-generated classes (blue), and the total image size distribution for the ten classes (red)

The ten blue bells on the graph correspond to the ten image classes. Note that the bells overlap, causing the resulting total distribution to be in good agreement with the original distribution of the training set. If we had ten separate networks, each for its own class, they would only generate images within their class boundaries and their sum would not be so similar to the original distribution.

When we compared the image size distributions for the training set and for the cGAN-generated sample, we found that the chi-square test statistic is 949 while the critical value corresponding for a 5% significance level with 100 degrees of freedom is 124,34. Although the chi-square test still shows that the difference is significant, the value of this criterion has decreased five times compared to the GAN results, and the resulting total distribution is much closer to the input one than the distribution for the classical GAN.

We checked the quality of the cGAN-generated images with a third-party software that is used for classification in the TAIGA-IACT project [19] and that determines the probability that an image is a gamma event image. The check showed that over 98% of the generated gamma images are recognized as valid gamma images with a probability of more than 90%. About 1% of the images are recognized as valid gamma images with a probability of 50% to 90%. And another 1% of the images have a probability below 50%, which means that these images are misinterpreted as the non-gamma images. These results turned out to be even better than the results of our classical GAN, for which the percentage of correct recognition was about 95% [10]. The examples of the generated images are shown in Figure 10.

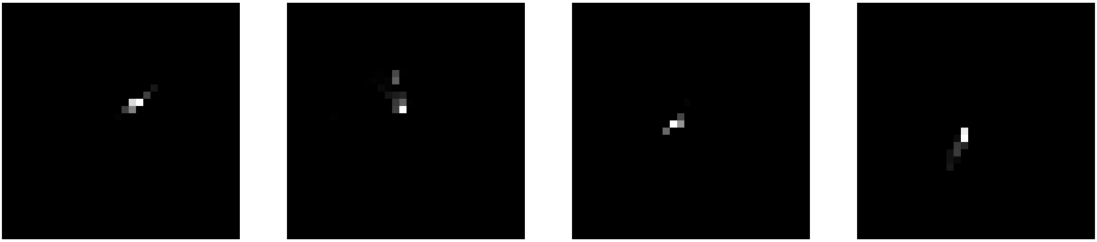

**Figure 10.** Examples of the gamma images generated by the cGAN

The generated images can be easily converted back to a hexagonal form. For example, it takes 3 seconds to back convert a batch of 10,000 images.

## 4. Conclusion

Summarizing the above, we can conclude that a conditional generative adversarial network generates images for the TAIGA-IACT experiment with a very good degree of accuracy with over 98% of the generated gamma images being recognized as valid ones by the third-party software.

Our cGAN generates images with a size value that falls within the boundaries of the requested class with a high degree of probability. Most of the rest images fall within the boundaries of the two nearest neighboring classes.

Compared to a classical GAN, the cGAN generates an output sample of images with an image size distribution that is much closer to that of the training set. The chi-square test statistic still shows that the difference is significant, but the value of this criterion has decreased five times compared to the GAN results.

The rate of image generation using cGAN is very high and is similar to the rate of image generation using GAN.

The results obtained will be useful for data augmentation and more accurate generation of realistic synthetic images similar to the ones taken by Imaging Atmospheric Cherenkov Telescopes.

## 5. Acknowledgments

The work was supported by the Russian Science Foundation, grant No. 22-21-00442. The authors would like to thank the TAIGA collaboration and, in particular, dr. E. B. Postnikov for providing model images that we used as the training set.